\documentstyle[11pt,newpasp,twoside,epsf]{article}
\def\edcomment#1{\iffalse\marginpar{\raggedright\sl#1\/}\else\relax\fi}
\reversemarginpar

\begin{document}
\title{New Grids of ATLAS9 Model Atmospheres}
\author{Fiorella Castelli}
\affil{Istituto di Astrofisica Spaziale e Fisica Cosmica, CNR, Via del Fosso  
del Cavaliere, 00133, Roma, Italy}
\affil{Osservatorio Astronomico di Trieste, Via G.B. Tiepolo~11, 
34131, Trieste, Italy}
\author{Robert L. Kurucz}
\affil{Harvard-Smithsonian Center for Astrophysics, 60 Garden Street, Cambridge, MA 02138, USA}

\begin{abstract}
New opacity distribution functions (ODFs) for several metallicities have been 
computed. The main
improvements upon previous ODFs computed by Kurucz (1990) are: (1) the 
replacement of the solar
abundances from Anders \& Grevesse (1989) with those from Grevesse \& Sauval
(1998); 
(2) the replacement of the TiO lines provided by Kurucz (1993) with the TiO lines from Schwenke (1998), as
distributed by Kurucz (1999a); (3) the addition of the H$_{2}$O lines 
from Partridge \& Schwenke~(1997), as
distributed by Kurucz (1999b); (4) the addition of the H~I-H~I and H~I-H$^{+}$
 quasi-molecular  absorptions
near 1600~\AA\ and 1400~\AA\ computed according to Allard et al. (1998). 
Other minor improvements are
related with some changes in a few atomic and molecular data. New grids 
of ATLAS9 model atmospheres for
T$_{\rm eff}$ from 3500~K to 50000~K and $\log$~g from 0.0~dex to 5.0~dex have
 been computed for several metallicities 
with the new ODFs. Preliminary comparisons of results from the old and new models have shown  differences
in the energy distributions of stars cooler than 4500~K, in the ultraviolet energy distribution of metal-poor A-type stars,
in the U$-$B and u$-$b color indices for T$_{\rm eff}$$\le$6750~K and in all
 the color indices for T$_{\rm eff}$$\le$4000~K. 
\end{abstract}

\section{Introduction}

Starting from 1992 Kurucz published new extended grids of model atmospheres 
for several metallicities [M/H] and
several microturbulent velocities $\xi$. In each grid, the models cover a 
range in T$_{\rm eff}$ and $\log$g from
3500~K to 50000~K and from 0.0~dex to 5.0~dex, respectively. All the models were computed with the ATLAS9 code,
which handles line opacity with the opacity distribution function method (ODFs).

Later on, Castelli extended Kurucz's models by computing some more grids for metallicities
with enhanced $\alpha$~elements abundances ($+$0.4~dex) and for $\xi$ values different from those
adopted by Kurucz. Furthermore, she recomputed the cool models in a few Kurucz grids by assuming the overshooting
option for the convection switched off (Castelli, Gratton \& Kurucz, 1997).

Both Kurucz and Castelli models were computed by using Kurucz's (1990) ODFs.  
They are all based on the solar abundances from Anders \& Grevesse (1989),
except for iron which is not the same in all the ODFs. In fact, the iron
abundance $\log$(N$_{Fe}$/N$_{tot}$) is $-$4.53~dex from Holweger et al. (1995) when ODFs for 
enhanced $\alpha$ elements abundances were computed and it is $-$4.37~dex 
from Anders \& Grevesse (1989) when ODFs for no enhanced $\alpha$ elements 
were generated. Therefore, it is not straightforward to state the net effect 
of the $\alpha$ elements abundances on the model structure.

Finally, the number of models is different in the various grids and also
in some grids some models have 72 and some others have 64 plane parallel 
layers. These differences may give problems when interpolations in the models
and in the grids are performed.

\section{The Aim of this Work}

The aim of this work is the computation of grids of homogenous ATLAS9 models,
where ``homogenous" means that:

\begin{description}
\item[(1)] The number of models is 476 in all the grids. Table~1 shows
which are the models.
\item[(2)] All the models have the same number of 72 plane parallel 
layers from $\log$$\tau_{Ross}$=$-$6.875 to $+$2.00 at steps of
$\Delta$$\log$$\tau_{Ross}$=0.125.
\item[(3)] All the models are computed with the updated solar abundances from
Grevesse \& Sauval (1998) and therefore with updated ODFs, which now also 
include H$_{2}$O lines.
\item[(4)] All the models are computed with the convection option switched on 
and with the overshooting option switched off. Mixing-length convection with 
l/H$_{p}$=1.25 is assumed for all the models. The convective flux decreases 
with increasing T$_{\rm eff}$ and it naturally disappears for T$_{\rm eff}$
of the order of 9000~K. 
\end{description}

\section{The new Opacity Distribution Functions}

The line opacity is included in the new ATLAS9 models through 
new-ODFs based on the Grevesse \& Sauval (1998) (GS98) solar abundances. 
Table~2 lists the differences
between GS98 abundances and those from Anders \& Grevesse (1989) which were
used in the Kurucz (1990) ODFs.

New-ODFs are computed for 57 values of temperature T (from 1995~K to 199526~K,
i.e. for $\log$T ranging from 3.30 to 5.30 at steps of 0.02~dex)
and 25 values of gas pressure $\log$P$_{g}$ (from $-$4.0 to +8.0 at steps of 0.5~dex),
while Kurucz (1990) ODFs were computed for 56 values of T (from 2089~K to
199526~K, i.e. for $\log$T ranging from 3.32 to 5.30 at steps of 0.02~dex) 
and 21 values of $\log$P$_{g}$ (from $-$2.0 to +8.0 at steps of 0.5~dex).
The larger ODF table should allow better modelling  the upper layers 
of cool stars and of giant stars.  

\begin{table}
\caption{The 476 models of each grid.}
\begin{tabular}{rcccccccccccr}
\tableline
& 0.0 & 0.5 & 1.0 & 1.5 & 2.0 & 2.5 & 3.0 & 3.5 & 4.0 & 4.5 & 5.0 & No of\\
&     &     &     &     &     &     &     &     &     &     &     &models\\
\tableline
3500& X & X & X & X & X & X & X & X & X & X & X & 11\\
3750& X & X & X & X & X & X & X & X & X & X & X & 11\\
4000& X & X & X & X & X & X & X & X & X & X & X & 11\\
4250& X & X & X & X & X & X & X & X & X & X & X & 11\\
4500& X & X & X & X & X & X & X & X & X & X & X & 11\\
4750& X & X & X & X & X & X & X & X & X & X & X & 11\\
5000& X & X & X & X & X & X & X & X & X & X & X & 11\\
5250& X & X & X & X & X & X & X & X & X & X & X & 11\\
5500& X & X & X & X & X & X & X & X & X & X & X & 11\\
5750& X & X & X & X & X & X & X & X & X & X & X & 11\\
6000& X & X & X & X & X & X & X & X & X & X & X & 11\\
6250&   & X & X & X & X & X & X & X & X & X & X & 10\\
6500&   & X & X & X & X & X & X & X & X & X & X & 10\\
6750&   & X & X & X & X & X & X & X & X & X & X & 10\\
7000&   & X & X & X & X & X & X & X & X & X & X & 10\\
7250&   & X & X & X & X & X & X & X & X & X & X & 10\\
7500&   & X & X & X & X & X & X & X & X & X & X & 10\\
7750&   &   & X & X & X & X & X & X & X & X & X &  9\\
8000&   &   & X & X & X & X & X & X & X & X & X &  9\\
8250&   &   & X & X & X & X & X & X & X & X & X &  9\\
8500&   &   &   & X & X & X & X & X & X & X & X &  8\\
8750&   &   &   & X & X & X & X & X & X & X & X &  8\\
9000&   &   &   & X & X & X & X & X & X & X & X &  8\\
9250&   &   &   &   & X & X & X & X & X & X & X &  7\\
9500&   &   &   &   & X & X & X & X & X & X & X &  7\\
9750&   &   &   &   & X & X & X & X & X & X & X &  7\\
10000&   &   &   &   & X & X & X & X & X & X & X &  7\\
10250&   &   &   &   & X & X & X & X & X & X & X &  7\\
10500&   &   &   &   & X & X & X & X & X & X & X &  7\\
10750&   &   &   &   & X & X & X & X & X & X & X &  7\\
11000&   &   &   &   & X & X & X & X & X & X & X &  7\\
11250&   &   &   &   & X & X & X & X & X & X & X &  7\\
11500&   &   &   &   & X & X & X & X & X & X & X &  7\\
11750&   &   &   &   & X & X & X & X & X & X & X &  7\\
12000&   &   &   &   &   & X & X & X & X & X & X &  6\\
12250&   &   &   &   &   & X & X & X & X & X & X &  6\\
12500&   &   &   &   &   & X & X & X & X & X & X &  6\\
12750&   &   &   &   &   & X & X & X & X & X & X &  6\\
13000&   &   &   &   &   & X & X & X & X & X & X &  6\\
14000&   &   &   &   & X & X & X & X & X & X & X &  7\\
15000&   &   &   &   &   & X & X & X & X & X & X &  6\\
16000&   &   &   &   &   & X & X & X & X & X & X &  6\\
17000&   &   &   &   &   & X & X & X & X & X & X &  6\\
18000&   &   &   &   &   & X & X & X & X & X & X &  6\\
19000&   &   &   &   &   & X & X & X & X & X & X &  6\\
20000&   &   &   &   &   &   & X & X & X & X & X &  5\\
21000&   &   &   &   &   &   & X & X & X & X & X &  5\\
22000&   &   &   &   &   &   & X & X & X & X & X &  5\\
23000&   &   &   &   &   &   & X & X & X & X & X &  5\\
\tableline
\tableline
\end{tabular}
\end{table}
\clearpage

\addtocounter{table}{-1}            

\begin{table}
\caption{cont.}
\begin{tabular}{rcccccccccccr}
\tableline
& 0.0 & 0.5 & 1.0 & 1.5 & 2.0 & 2.5 & 3.0 & 3.5 & 4.0 & 4.5 & 5.0 & No of\\
&     &     &     &     &     &     &     &     &     &     &     &models\\
\tableline
24000&   &   &   &   &   &   & X & X & X & X & X &  5\\
25000&   &   &   &   &   &   & X & X & X & X & X &  5\\
26000&   &   &   &   &   &   & X & X & X & X & X &  5\\
27000&   &   &   &   &   &   &   & X & X & X & X &  4\\
28000&   &   &   &   &   &   &   & X & X & X & X &  4\\
29000&   &   &   &   &   &   &   & X & X & X & X &  4\\
30000&   &   &   &   &   &   &   & X & X & X & X &  4\\
31000&   &   &   &   &   &   &   & X & X & X & X &  4\\
32000&   &   &   &   &   &   &   &   & X & X & X &  3\\
33000&   &   &   &   &   &   &   &   & X & X & X &  3\\
34000&   &   &   &   &   &   &   &   & X & X & X &  3\\
35000&   &   &   &   &   &   &   &   & X & X & X &  3\\
36000&   &   &   &   &   &   &   &   & X & X & X &  3\\
37000&   &   &   &   &   &   &   &   & X & X & X &  3\\
38000&   &   &   &   &   &   &   &   & X & X & X &  3\\
39000&   &   &   &   &   &   &   &   & X & X & X &  3\\
40000&   &   &   &   &   &   &   &   &   & X & X &  2\\
41000&   &   &   &   &   &   &   &   &   & X & X &  2\\
42000&   &   &   &   &   &   &   &   &   & X & X &  2\\
43000&   &   &   &   &   &   &   &   &   & X & X &  2\\
44000&   &   &   &   &   &   &   &   &   & X & X &  2\\
45000&   &   &   &   &   &   &   &   &   & X & X &  2\\
46000&   &   &   &   &   &   &   &   &   & X & X &  2\\
47000&   &   &   &   &   &   &   &   &   & X & X &  2\\
48000&   &   &   &   &   &   &   &   &   & X & X &  2\\
49000&   &   &   &   &   &   &   &   &   & X & X &  2\\
50000&   &   &   &   &   &   &   &   &   &   & X &  1\\
\tableline
\tableline
\end{tabular}
\end{table}

\begin{table}
\caption{Solar abundances different in Anders \& Grevesse (1989) (AG89) and in
Grevesse \& Sauval (1998) (GS98). The abundances are given as 
$\log$(N$_{elem}$/H$_{H}$).}
\begin{tabular}{rcccccccccccccccccc}
\tableline
&   He & Li & Be & B & C & N & O & Ne & S \\
\tableline
AG89&$-$1.01&$-$10.84&$-$10.85&$-$9.40&$-$3.44&$-$3.95&$-$3.07&$-$3.91&$-$4.79\\
GS98&$-$1.07&$-$10.90&$-$10.60&$-$9.45&$-$3.48&$-$4.08&$-$3.17&$-$3.92&$-$4.67\\
\\
\tableline
 & Ar & Sc &Ti & Fe &Se & Kr& Sr& Cd& Xe \\
\tableline
AG89&$-$5.44&$-$8.90&$-$7.01&$-$4.33&$-$8.65&$-$8.77&$-$9.10&$-$10.14&$-$9.77\\
GS98&$-$5.60&$-$8.83&$-$6.98&$-$4.50&$-$8.59&$-$8.69&$-$9.03&$-$10.23&$-$9.83\\
\tableline
\tableline
\end{tabular}
\end{table}

Except for TiO, new-ODFs are computed with the atomic and molecular line
lists as in the Kurucz (1990) ODFs, but some molecular lists have been
extended to more bands (i.e. CN, OH, and SiO) and some bugs 
in the line lists have been corrected.

In new-ODFs, the TiO lines computed by Kurucz (1993) with semi-empirical 
methods have been replaced by the TiO lines from Schwenke (1998), as distributed
by Kurucz (1999a) on the CD-ROM No~24. Furthermore, the H$_{2}$O lines from
Partridge \& Schwenke (1997), as distributed by Kurucz (1999b) on the 
CD-ROM No~26, were added to the other lines.
New-ODFs also include H~I-H$^{+}$ and H~I-H~I quasi-molecular absorptions at
1400~\AA\ and 1600~\AA, respectively, due to the collision of neutral 
hydrogen in the ground-state with ionized hydrogen and other neutral hydrogen
atoms (Castelli \& Kurucz, 2001). These absorptions are observed as broad
features in $\lambda$~Boo stars and in metal-poor A-type stars. 

The treatment of the overlapping lines at the end of the term series was
also improved in the new-ODFs.

New-ODFs are computed on an Alpha Digital Personal Workstation running
OPEN VMS using the codes XNFDF and DFSYNTHE. The XNFDF code pretabulates number
densities and continua opacities. The DFSYNTHE code computes a line opacity
spectrum for each T,P$_{gas}$ couple from 89.7~\AA\ to 100000~\AA\ at
500000 resolution. Line opacities are then sorted in the same wavelength
intervals as in the Kurucz (1990) ODFs.

\begin{figure}
%\plotone
\plotfiddle{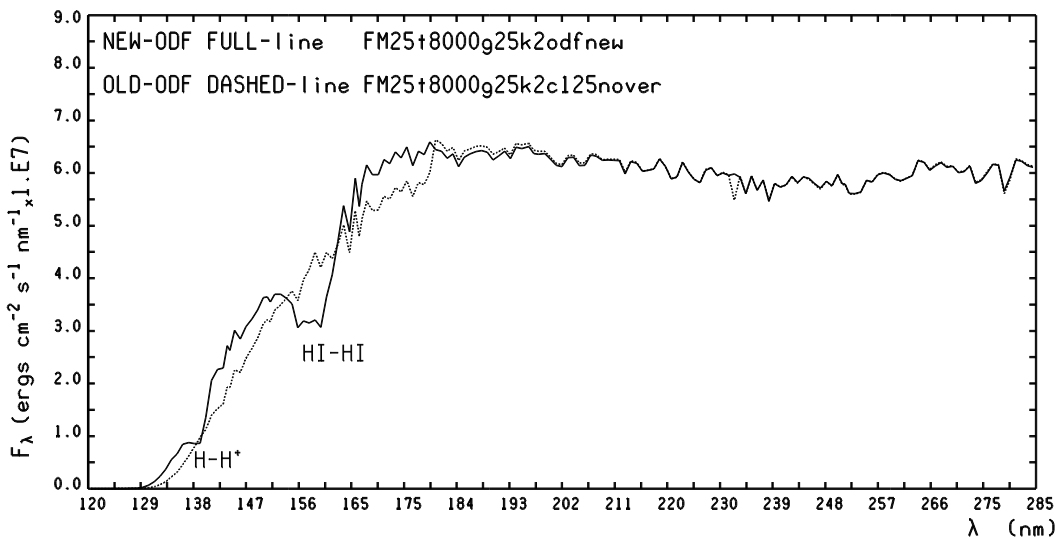}{6.5cm}{0}{70}{70}{-200pt}{-10pt}
\caption{Comparison of the ultraviolet flux from the old-ODF model (thin dashed line)
and the new-ODF model (thick full line) with parameters  
T$_{\rm eff}$=8000~K, $\log$g=2.5, [M/H]=$-$2.5, $\xi$=2.0~km~s$^{-1}$}
\end{figure}

\begin{figure}
%\plotone
\plotfiddle{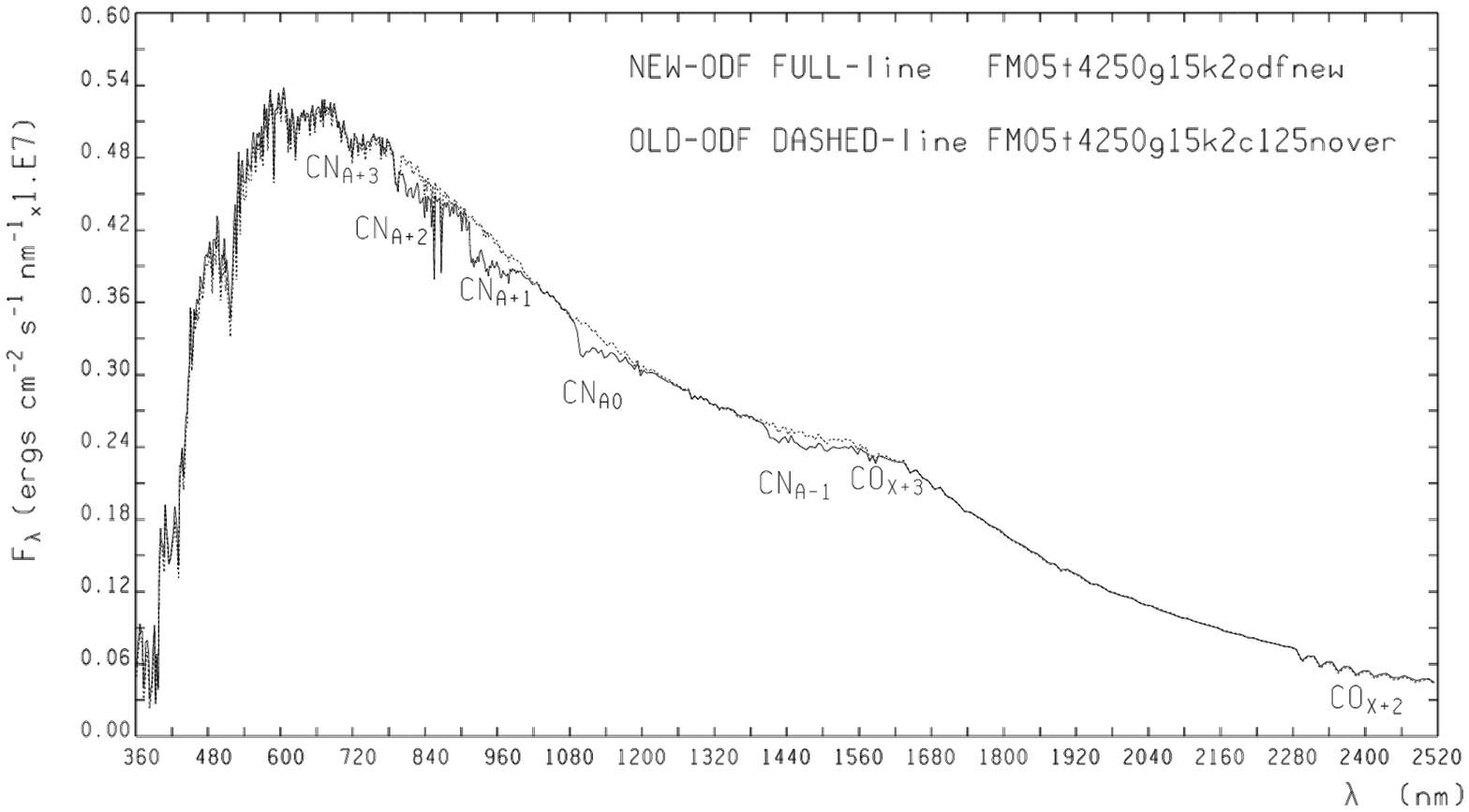}{6.5cm}{0}{70}{70}{-200pt}{-15pt}
\caption{Comparison of fluxes from the old-ODF model (thin dashed line)
and the new-ODF model (thick full line) with 
parameters T$_{\rm eff}$=4250~K, $\log$g=1.5, [M/H]=$-$0.5, 
$\xi$=2.0~km~s$^{-1}$.  
The bands of CN and CO are indicated in the figure.}
\end{figure}

\begin{figure}
%\plotone
\plotfiddle{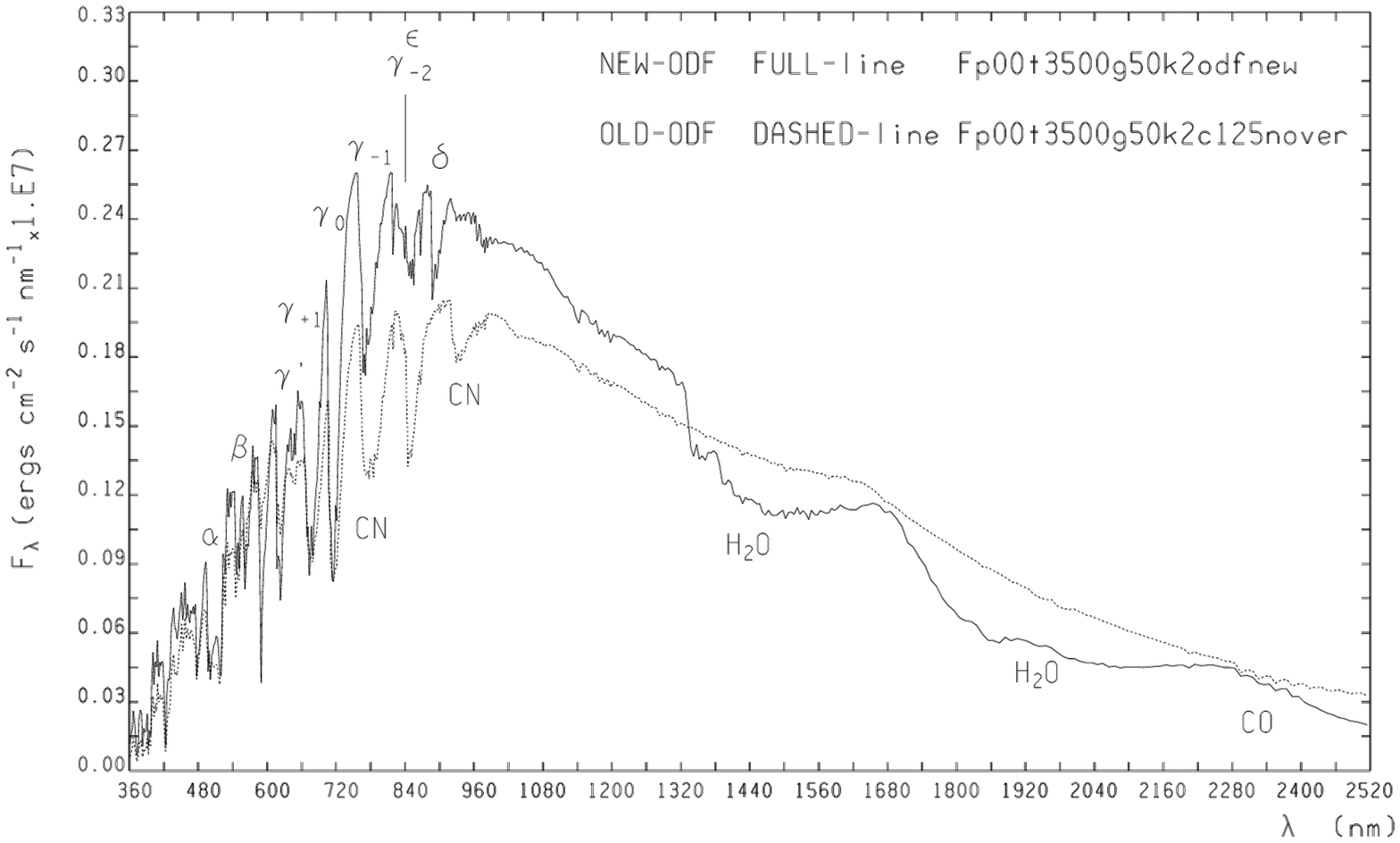}{7.0cm}{0}{70}{70}{-200pt}{-15pt}
\caption{Comparison of fluxes from the old-ODF model (thin dashed line) 
and the new-ODF model (thick full line) with parameters 
T$_{\rm eff}$=3500~K, $\log$g=5.0, [M/H]=0.0, $\xi$=2.0~km~s$^{-1}$.
The bands of TiO, CN, H$_{2}$O, and CO are indicated in the figure.}
\end{figure}

\begin{figure}
%\plotone
\plotfiddle{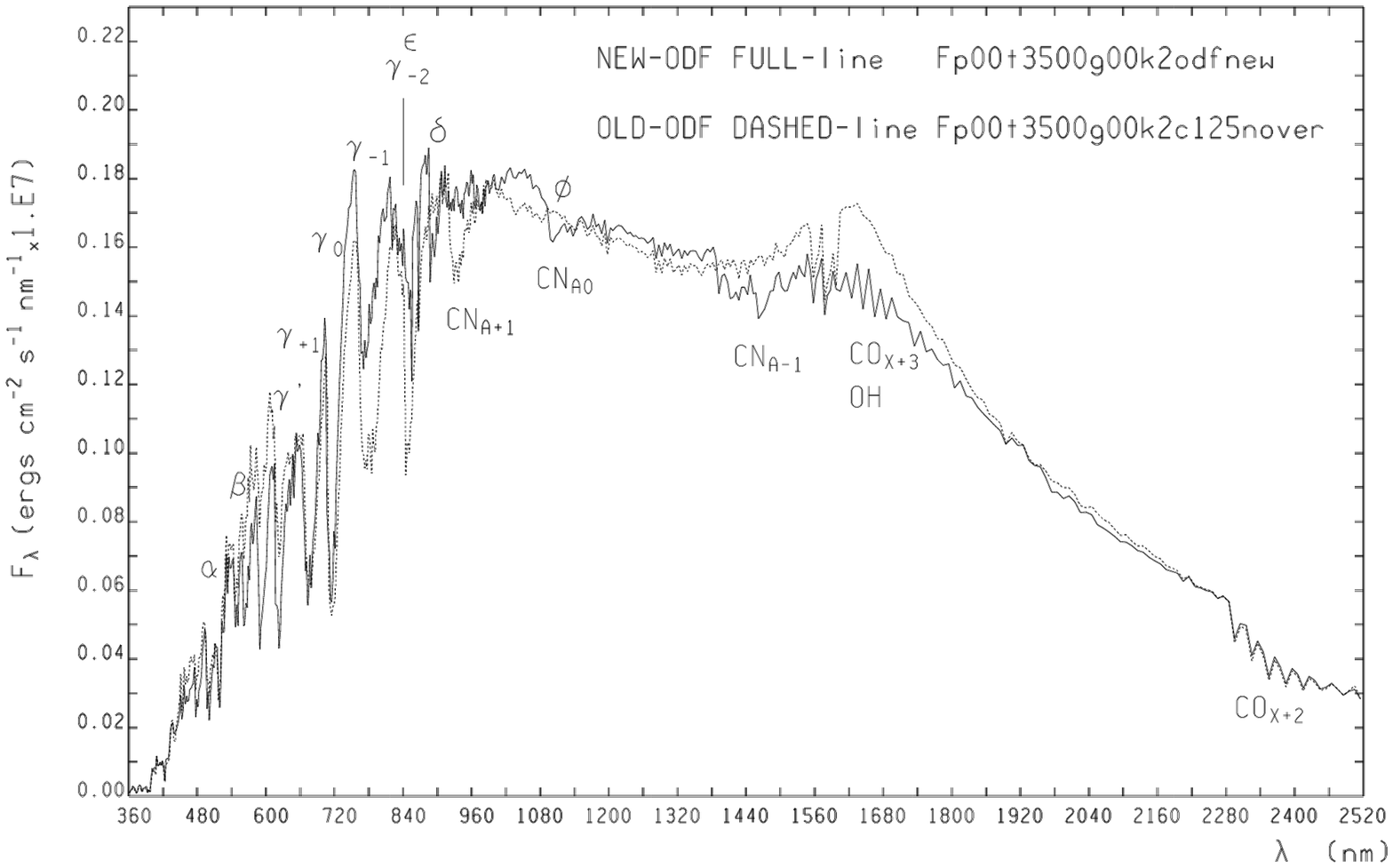}{7.2cm}{0}{70}{70}{-200pt}{-15pt}
\caption{Comparison of fluxes from the old-ODF model (thin dashed line) 
and the new-ODF model (thick full line) with parameters 
T$_{\rm eff}$=3500~K, $\log$g=0.0, [M/H]=0.0, $\xi$=2.0~km~s$^{-1}$.
The bands of TiO, CN, OH, and CO are indicated in the figure.}
\end{figure}

\begin{figure}
%\plotone
\plotfiddle{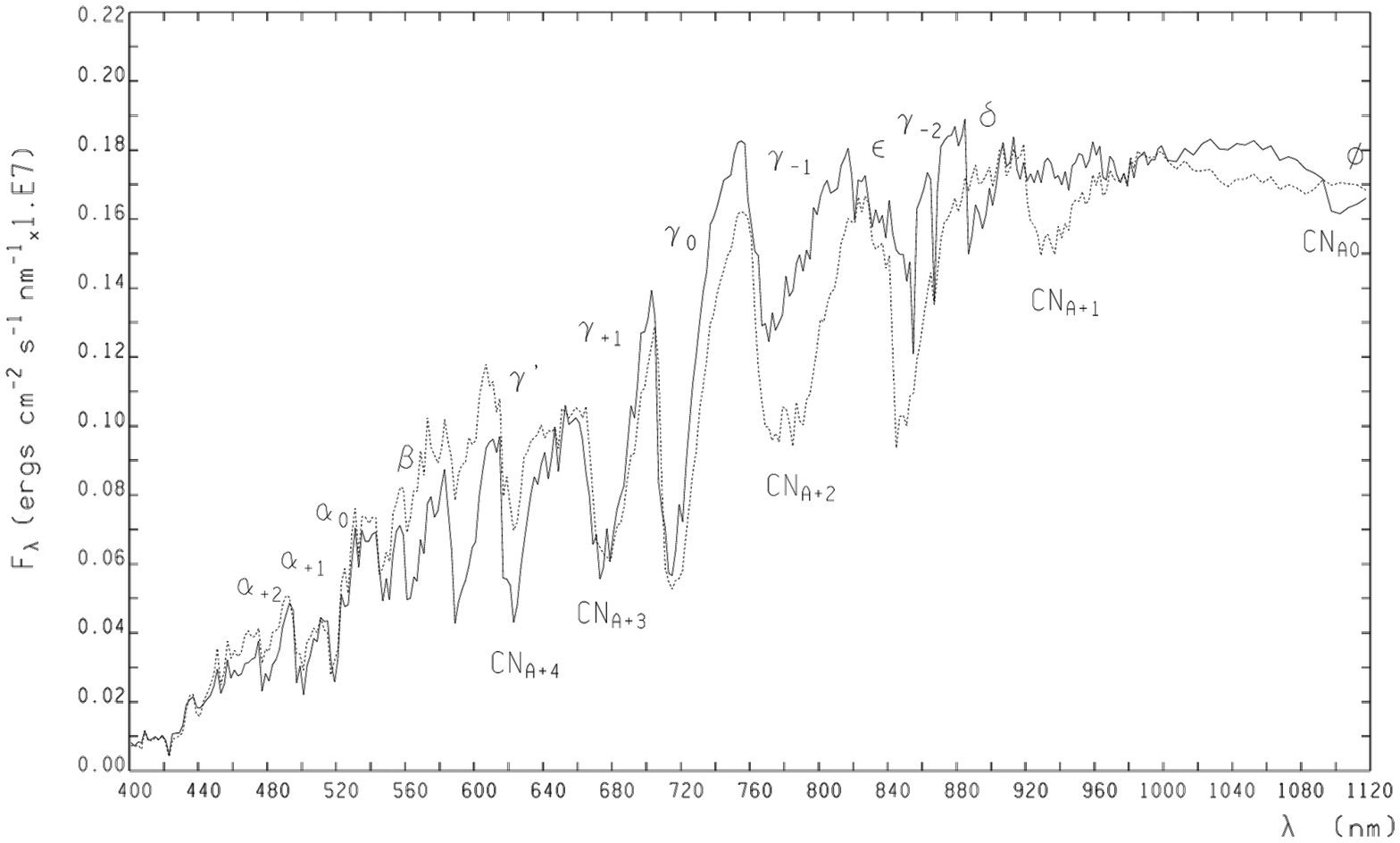}{6.5cm}{0}{70}{70}{-200pt}{-15pt}
\caption{The same as Fig.~4, but for the smaller wavelength region 400-1120~nm
plotted on a larger wavelength scale.}
\end{figure}

\section{The New Grids and Future Work}

New full grids of model atmospheres computed up to now are those for:

[M/H]=0.0, $-$0.5, $-$0.5a, $-$1.0, $-$1.5 and $\xi$=2~km~s$^{-1}$

where ``a'' means that the abundances of the $\alpha$ elements 
O, Ne, Mg, Si, S, Ar, Ca, and Ti are enhanced by $+$0.4~dex over the solar
or solar-scaled abundances.

Together with the models we have computed corresponding grids of fluxes and
color indices in the UBV and RIJKL Johnson system, in the VRI Cousins system,
and in the uvby Str\"omgren system.
The new grids are available at http://kurucz.harvard.edu/grids/gridxxxodfnew.

Small new-ODF grids for a few values of T$_{\rm eff}$ and $\log$g have been
also computed for [M/H]= $-$1.5a, $-$2.0, $-$2.0a, $-$2.5, $-$2.5a, and
$-$3.0a (Castelli \& Cacciari, 2001). We plan to extend these small grids to
all the models listed in Table~1. In addition to models and fluxes computed
for $\xi$=2~km~s$^{-1}$ we plan to compute models and fluxes for
$\xi$=0, 1, 4, and 8 km~s$^{-1}$.
Color indices in several photometric systems will be also computed.

\begin{figure}
%\plotone
\plotfiddle{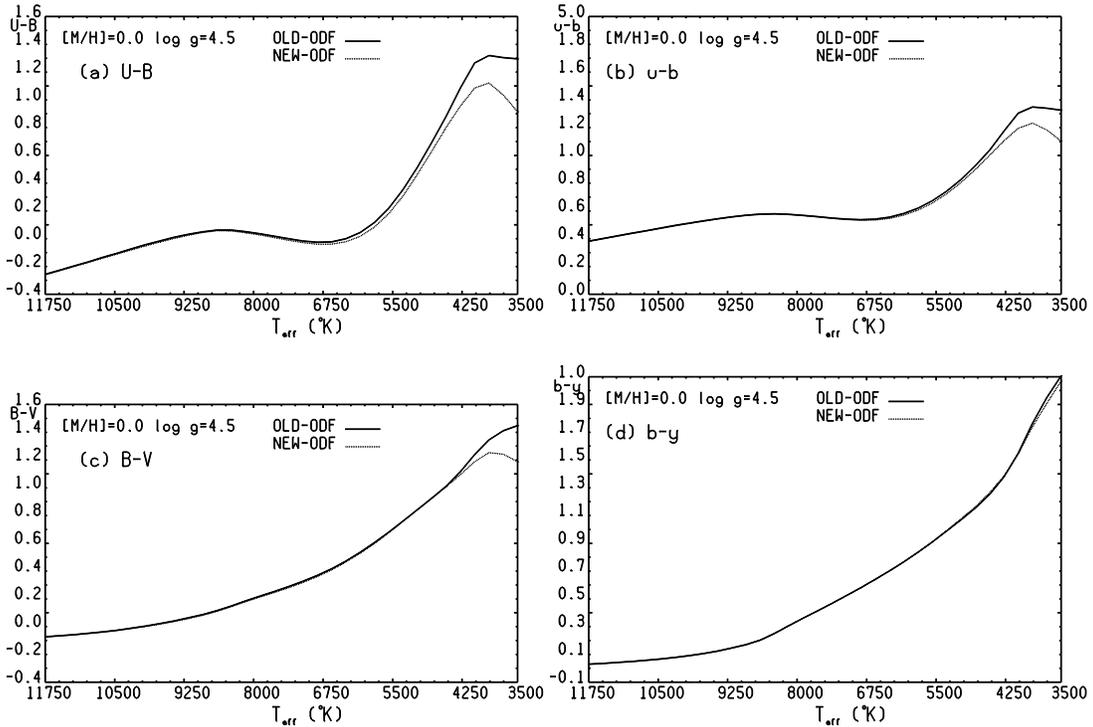}{9.0cm}{0}{70}{70}{-225pt}{-150pt}
\caption{ T$_{\rm eff}$-color relations for $\log$g=4.5 and [M/H]=0.0
when old-ODF models (thick line) and new-ODF models (thin line) 
are used: (a) for U$-$B;
(b) for u$-$b; (c) for B$-$V; (d) for b$-$y.}  
\end{figure}

\section{Comparison between Old and New Models}

Preliminary comparisons of energy distributions have shown that the 
different solar abundances do not affect the models very much. In fact,
the differences between old-ODF and new-ODF models are significant only
when additional absorptions introduced in the new-ODFs play an important role.

Remarkable differences which increase with decreasing gravity and/or 
metallicity can be observed in the 1250-2000~\AA\ region for T$_{\rm eff}$
between 9000~K and 7000~K. They are mostly due to the H~I-H$^{+}$ and H~I-H~I
quasi-molecular absorptions. 
Fig.~1 shows the energy distributions from old-ODF
and new-ODF models for T$_{\rm eff}$=8000~K, $\log$g=2.5, [M/H]=$-$2.5,
$\xi$=2~km~s$^{-1}$. 

The new TiO line list, the addition of H$_{2}$O lines and some improvements
in the molecular data have greatly modified the energy distributions of 
models cooler than 4500~K. Examples are given in Figs. 2, 3, 4, and 5.
Figure~5 shows in more detail
 the TiO and CN absorptions in the range 4000-11200~\AA\
for the model T$_{\rm eff}$=3500~K, $\log$g=0.0, [M/H]=0.0, $\xi$=2~km~s$^{-1}$.

Color indices U-B and u-b from new-ODF models are redder than the indices from
the old-ODF models for T$_{\rm eff}$$<$6500~K (Fig.~6a,b). The difference
increases with decreasing T$_{\rm eff}$. For instance, the difference for 
U$-$B amounts to 75~K when the
model T$_{\rm eff}$=5750~K, $\log$g=4.5, [M/H]=0.0, $\xi$=2~km~s$^{-1}$ 
is considered.

Both (B-V) and (b-y) from new-ODFs are redder than those from old-ODFs for
T$_{\rm eff}$$<$4500~K, but the differences for (b-y) are almost negligible
 (Fig.~6c,d).


\begin{references}

\reference Allard, N.F., Drira, I., Gerbaldi, M., Kielkopf, J., 
\& Spieldfield, A.,1998, A\&A, 335, 1124  

\reference Anders, A., \& Grevesse, N., 1989, Geochim. Cosmochim. Acta, 53, 197

\reference Castelli, F., \& Cacciari, C., 2001, A\&A, 380, 630

\reference Castelli, F., Gratton, R. G., \& Kurucz, R. L., 1997, A\&A, 318, 841

\reference Castelli, F., \& Kurucz, R. L., 2001, A\&A, 372, 260

\reference Grevesse, N., \& Sauval, A. J., 1998, Space Sci. Rev., 85, 161

\reference Holweger, H., Kock, M., \& Bard, A., 1995, A\&A, 296, 233

\reference Kurucz, R. L., 1990, ``Stellar Atmospheres: Beyond Classical Models",
NATO Asi Ser., ed. L. Crivellari et al.,441

\reference Kurucz, R. L.,1993,``Diatomic Molecular Data for Opacity Calculations",
CD-ROM No 15

\reference Kurucz, R. L., 1999a, ``TiO Linelist from Schwenke (1998)",
 CD-ROM No 24

\reference Kurucz, R. L., 1999b, ``H$_{2}$O Linelist from Partridge and 
Schwenke (1997), no IDs", CD-ROM No 26

\reference Partridge, H., \& Schwenke, D. W., 1997, J. Chem. Phys., 106, 4618

\reference Schwenke, D. W., 1998, Faraday Discuss., 109, 321 

\end{references}
\end{document}